\documentclass[fleqn,twoside]{article}

\newcommand{\AmS}{{\protect\the\textfont2  A\kern-.1667em\lower.5ex\hbox{M}\kern-.125emS}}

\usepackage{espcrc2}

\usepackage{graphicx}
\usepackage[figuresright]{rotating}

\title{Deep Inelastic Scattering in Holographic AdS/QCD Models}

\author{ C. A.  Ballon Bayona\address[CBPF]
{Centro Brasileiro de Pesquisas F\'{i}sicas, \\
Rua Dr. Xavier Sigaud 150, Urca, 22290-180 Rio de Janeiro, RJ, Brazil\\}
 \thanks{email: ballon@cbpf.br},
        Henrique Boschi-Filho\address[UFRJ]{Instituto de F\'{\i}sica,
Universidade Federal do Rio de Janeiro, \\
Caixa Postal 68528, 21941-972 Rio de Janeiro, RJ, Brazil}\thanks{email: boschi@if.ufrj.br },
Nelson R.F. Braga\addressmark[UFRJ] \thanks{email: braga@if.ufrj.br}}
       
\begin{document}

\begin{abstract} 
We review the description of deep inelastic scattering using some AdS/QCD phenomenological models.  
\vspace{1pc}
\end{abstract}

\maketitle

\section{ Introduction }

Important non perturbative aspects of strong interactions can be studied using the so called 
AdS/QCD phenomenological models. These models are inspired in the AdS/CFT correspondence \cite{Maldacena:1997re,Gubser:1998bc,Witten:1998qj}, 
which is a duality between string theory and superconformal gauge theories at large 't Hooft coupling. 
The form of this correspondence that is more relevant in studying hadronic physics, 
relates string theory in $AdS_5\times S^5 $ space to a superconformal gauge theory on the four dimensional boundary.

One of the simplest AdS/QCD models, the hard wall model, consists in the introduction of a hard cut off in anti-de Sitter (AdS) space in order to break conformal invariance. The position of the cut off represents an infrared mass scale for the gauge theory. This model makes it possible to reproduce the scaling of hadronic elastic scattering amplitudes at fixed angles from string theory in AdS space \cite{Polchinski:2001tt}.
String theory predictions in flat space were in contrast to the experimentally observed behavior of hadronic scattering at high energies.
This scaling was also obtained \cite{BoschiFilho:2002zs} from a map between bulk and boundary quantum
states in AdS space with a cut off.

It was also shown that it is possible to calculate hadronic masses using the AdS/QCD hard wall model. 
In the AdS/CFT correspondence, normalized solutions for fields in AdS space are dual
to states in the dual boundary gauge theory. Using this duality, for the case of glueballs, 
and the idea of introducing boundary conditions in the AdS infrared cut off, the spectrum of scalar glueballs
was found \cite{BoschiFilho:2002ta,BoschiFilho:2002vd}.  
This approach made it possible to find also the glueball Regge trajectories and compare them with the Pomeron trajectories \cite{BoschiFilho:2005yh}. 

An alternative AdS/QCD model consists of a space with AdS metric where there is a constant non uniform 
dilaton background field. In this model, called soft wall, the dilaton acts effectively as a smooth  
infrared cut off and leads to linear Regge trajectories for mesons \cite{Karch:2006pv}.  
The soft wall was also used to calculate  glueball masses and decay constants \cite{Colangelo:2007pt,Forkel:2007ru}. For a recent review on glueballs see, {\it e. g.}, 
\cite{Mathieu:2008me}.

In the original AdS/CFT correspondence there are only fields in the adjoint representation of the $SU(N_c)$ 
gauge group. This happens because, in the case of $AdS_5\times S^5 $ space, there is a set of 
$N_c$ colored  D3 branes and the boundary fields show up as excitations of open strings attached 
to these branes. 
In order to introduce matter fields, that are in the fundamental representation, like quarks, one can 
add $N_f$ D7 probe branes in the space \cite{Karch:2002sh,Kruczenski:2003be,Hong:2003jm}.
In this D3-D7 brane model, open strings with an endpoint  on a D3 and the other on a D7 brane 
have color and flavor. So, their excitations can include quark like fields.  
In this model, mesons are described by strings with both endpoints on D7 branes, corresponding
to D7 brane fluctuations. Masses for mesons in this model were calculated in 
\cite{Kruczenski:2003be}. 

The AdS/QCD models have been used not only to study the mass spectrum but also other 
important properties that involve non perturbative aspects of QCD, like the description of the hadronic structure.  
A very important process that gives us informations about the internal structure of hadrons 
is the deep inelastic scattering (DIS). It consists of a scattering of a highly energetic lepton 
off of a hadron. This process was investigated using the hard wall model in \cite{Polchinski:2002jw}. 
The matrix element of the four dimensional electromagnetic current of the hadron is mapped into 
a ten dimensional supergravity interaction action, that holographically describes the scattering process. This way, the structure functions are calculated.

The description of DIS using the soft wall AdS/QCD model was presented in \cite{BallonBayona:2007qr}.
The hadronic structure functions found from the hard and soft wall models are different, although
they agree at leading order. This is a non trivial result, since the calculation of the hadronic 
structure functions involve the mass spectrum of the final hadronic states, that are different in 
the soft and hard wall models. 

The inclusion of flavour degrees of freedom in a DIS process was considered in \cite{BallonBayona:2008zi}
by using the D3-D7 brane model. In this article two possibilities were considered for the current interacting with the hadron. 
One is the current living in the AdS bulk and the other a current living on the D7 brane, representing a flavour current. 
The structure functions obtained with the bulk current are similar to the hard and soft wall results but the case of the 
flavour current leads to different results. 

In these studies of the DIS using gauge string duality it was considered that the final hadronic state 
was represented by an operator with the same conformal dimension of the initial one. 
In ref. \cite{BallonBayona:2007rs} the case of final states with higher conformal dimensions was studied. 
The increase in the conformal dimension is interpreted as an increase in the number of hadronic constituents. 
Summing over all conformal dimensions allowed, a scaling similar to geometric scaling \cite{Stasto:2000er} was found. 

In this article we briefly review these results for the DIS in the hard wall, soft wall and D3-D7 models and also the case with final states with higher conformal dimension. For recent related studies of hadronic interactions from gauge/string duality see, for example, 
\cite{Grigoryan:2007vg,Grigoryan:2007my,Brodsky:2007hb,Hatta:2007he,Hatta:2007cs,Kwee:2007nq,Hatta:2008tx,Cornalba:2008sp,Hatta:2008tn,Pire:2008zf,Albacete:2008ze,Hatta:2008qx,RodriguezGomez:2008zp,Hatta:2008st,Gao:2009ze,Hatta:2009ra}.

\section{ Deep Inelastic Scattering in AdS/QCD models }

\begin{figure}
\setlength{\unitlength}{0.05in}
\vskip 2.2cm
\hskip 3.cm
\begin{picture}(0,0)(10,0)
\rm
\thicklines
\put(1,14.5){$\ell$}
\put(3,15){\line(2,-1){7}}
\put(3,15){\vector(2,-1){4}}
\put(18,14.5){$\ell$}
\put(17,15){\line(-2,-1){7}}
\put(10,11.5){\vector(2,1){4.2}}
\put(9.5,8){$q$}
\bezier{300}(10,11.5)(10.2,10.7)(11,10.5)
\bezier{300}(11,10.5)(11.8,10.3)(12,9.5)
\bezier{300}(12,9.5)(12.2,8.7)(13,8.5)
\bezier{300}(13,8.5)(13.8,8.3)(14,7.5)
\put(0,-2){$P$}
\put(3,0){\line(2,1){10.5}}
\put(3,0){\vector(2,1){6}}
\put(16,6){\circle{4.9}}
\put(27,-2){$X$}
\put(18.5,5.5){\line(3,-1){8}}
\put(18.3,5){\line(2,-1){8}}
\put(18,4.5){\line(3,-2){7.5}}
\put(17.5,3.8){\line(1,-1){6}}
\end{picture}
\parbox{4.1 in}{\caption{ Deep inelastic scattering.} }
\end{figure}
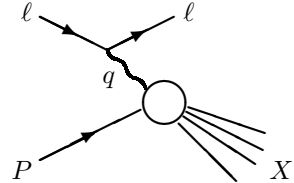

In deep inelastic scattering (DIS) a lepton scatters from a hadron of momentum $P^\mu $ through the exchange of a virtual photon of momentum $q^\mu$ (for a review see\cite{Manohar:1992tz}). 
When one measures the inclusive cross section, the final hadronic state $X$ with momentum $P_X^\mu$ is not observed. Then, the cross section involves a sum over all possible final states $X$.     
The kinematical variables used in general to describe DIS are $q^2$ and the Bjorken parameter $x \equiv  -q^2 /2P\cdot q \,$.

The DIS cross section can be calculated from the hadronic tensor:  
\begin{eqnarray}
W^{\mu\nu} \!\!\! &=& \!\!\! i \int d^4y\, e^{iq\cdot y} \langle P, {\cal Q} \vert \, \Big[ J^\mu (y) , J^\nu (0) \Big] 
\, \vert P, {\cal Q} \rangle \,,\nonumber\\
& &  \label{HadronicTensor}
\end{eqnarray}

\noindent where $ J^\mu(y)$ is the electromagnetic hadron current and $ {\cal Q} $ is the electric charge of the initial  hadron. Using Lorentz and gauge invariance, the hadronic tensor, for the spinless case,
has a tensor decomposition in terms of the structure functions $F_1 (x,q^2) $ and $F_2 (x,q^2) $  
\cite{Manohar:1992tz}
\begin{eqnarray}
W^{\mu\nu} \!\!\!\! &=& \!\! F_1 (x,q^2)  \Big( \eta^{\mu\nu} - \frac{q^\mu q^\nu}{q^2}  \Big) 
\nonumber\\ &+& \frac{2x}{q^2} F_2 (x,q^2)  \Big( P^\mu +  \frac{q^\mu}{2x}  \Big) 
\Big( P^\nu +  \frac{q^\nu}{2x}  \Big)
\, ,  \label{Structure}
\end{eqnarray}

\noindent where we use $\eta_{\mu\nu}={\rm diag}(-,+,+,+)$. 

The DIS cross section is related by the optical theorem  to 
the forward hadron-photon Compton scattering amplitude, described by the  tensor 
\begin{eqnarray}
T^{\mu\nu} \!\!\!&=& \!\!\!  i \int \! d^4y e^{iq\cdot y} \langle P, {\cal Q} \vert \, {\cal T} 
\Big(  J^\mu (y) J^\nu (0) \Big)  
 \vert P, {\cal Q} \rangle \,,\nonumber\\
& &
\label{forwardamplitude}
\end{eqnarray}

\noindent that can be decomposed in the same way as eq. (\ref{Structure}) but with 
structure functions ${\tilde F}_1 (x,q^2)  $ and ${\tilde F}_2 (x,q^2) $, related  
to the DIS structure functions by
\begin{equation}
\label{optical}
F_{1,2} (x,q^2) \equiv 2 \pi \,{\rm Im }\,{\tilde F}_{1,2} (x,q^2)\,.
\end{equation}

The imaginary part of the forward Compton scattering amplitude can be expressed in terms of a sum over the intermediate states $X$ with mass $M_X\,$ 
\begin{eqnarray}
\label{Imag}
& {\rm Im}& \!\!\! \!\!\!   T^{\mu\nu}\, = \, 2 \pi^2 \, \sum_X 
\,\delta \Big( M_X^2 + (P+q)^2 \, \Big)
\nonumber\\  
& \times & \!\! \!\!\! \!\!\!  \langle P, {\cal Q} \vert J^\nu ( 0 )   \vert P + q, X \rangle \langle P + q ,  X \vert J^\mu ( 0 )   \vert P, {\cal Q} \rangle\,.\nonumber\\
\end{eqnarray}

\noindent In this expression we see that there is a dependence on the masses $M_X$ of the final states. These masses are different in the models considered.   


\subsection{ DIS in the hard wall model}

The hard wall model consists of a space $AdS_5 \times W \,$, where $W $ is a five dimensional  compact space like $\,S^5\,$. The metric $g_{MN}$ is: 
\begin{eqnarray}
\label{AdS} 
ds^2 &\equiv& \!\!\!  g_{MN} \,dx^M dx^N \,= \, \frac{R^2}{z^2}( dz^2  +
\eta_{\mu\nu} dy^\mu dy^\nu  ) \nonumber\\ &+&   \,R^2 ds_W^2\,\,\,. 
\end{eqnarray}

\noindent We introduce a hard cut off, represented by the condition: $ 0 \le z \le 1/\Lambda \,$ , where $\Lambda\,$ works as an infrared cut off for the four dimensional boundary theory and can be 
interpreted as the QCD like scale.  $R$ is the AdS radius defined by $R^4=4\pi\, N_c\, g_s\, {\alpha'}^2$.  

The prescription relating the matrix elements of the boundary hadronic $U(1)$  current to string theory in
the bulk depends on the value of the Bjorken parameter $x$. 
For $ x >> 1/ \sqrt{gN_c} $  supergravity approximation can be used. For smaller values of $x$ 
one should include massive string states, as discussed in \cite{Polchinski:2002jw}. 
Here we will consider only this supergravity case. For a scalar hadron, the prescription relates 
the matrix elements of the current to a ten dimensional supergravity interaction action:
\begin{eqnarray}
(2\pi)^4\delta^4(P_X-P-q)\eta_\mu\langle P+q,X \vert J^\mu(0) 
\vert P,{\cal Q}\rangle\hskip -.5cm\cr
= i {\cal Q}\int d^{^{10}}x \sqrt{-g} A^{ \texttt m} \Big( \Phi_i\partial_{\texttt m} 
\Phi_X^\ast - \Phi_X^\ast  \partial_{\texttt m} \Phi_i  \Big),
\label{INTERACTION}
\end{eqnarray}

\noindent where $ \eta_\mu $ is the virtual photon polarization, $A_{\texttt m} (x)=(A_z,A_\mu)$ is a Kaluza-Klein gauge field, $ \Phi_i $ and $ \Phi_X$ are the fields representing the initial and final hadronic  states.  The hadronic mass spectrum in the hard wall implies that   
\begin{equation}
\label{nova3}
\sum_X \,\delta \Big( M_X^2 + (P+q)^2 \, \Big) 
\,=\, \frac{1}{2\pi s^{1/2} \,\Lambda\,}\,,
\end{equation}
\noindent with $s=-(P+q)^2$. The scalar structure functions in the hard wall model at leading order in $ \Lambda^2 /q^2 $, for
this supergravity case, take the form:
\begin{equation}
F_1 (x,q^2)\,=\,0 \,
\end{equation}
\begin{equation}
F_2 (x,q^2) = \pi C_0 {\cal Q}^2 \left( \frac{\Lambda^2}{q^2} 
\right)^{\Delta - 1} \!\!\!\!\! x^{\Delta + 1} (1- x)^{\Delta - 2},
\label{hwsf}
\end{equation}

\noindent where $C_0$ is a dimensionless normalization constant and  $\Delta$ is the scaling dimension of the scalar state. 


\subsection{ DIS in the soft wall model}
In the soft wall model\cite{Karch:2006pv} there is an $AdS_5$ space with a static dilaton 
background field $\varphi $ chosen as $\varphi = c z^2$. This dilaton acts as an infrared cut off. 
The constant $c$, with dimension of mass squared, is related to the QCD infrared scale.

In  \cite{BallonBayona:2007qr} we considered a ten dimensional extension of this model
and assumed the prescription for calculating the matrix element of the hadronic current in 
the supergravity regime:  
\begin{eqnarray}
(2 \pi)^4  \delta^4 ( P_X - P - q ) \,\eta_\mu \,  \langle P_X, X \vert J^\mu (0)
\vert P,{\cal Q} \rangle \cr
= i {\cal Q} \int d^{^{10}}x \sqrt{-g}\, e^{-\varphi} A^{\texttt m} 
\Big( \Phi_i\partial_{\texttt m} \Phi_X^\ast \cr 
- \Phi_X^\ast  \partial_{\texttt m} \Phi_i \Big). 
\label{INTERACTION2}
\end{eqnarray}
 
The mass spectrum of the final hadronic states implies that in the soft wall model  
\begin{equation}
\label{nova2}
\sum_X \,\delta \Big( M_X^2 + (P+q)^2 \, \Big) 
\,=\,  \frac{1}{4c}\, .
\end{equation}

Using these results,  we found the structure functions for scalar hadrons: 
\begin{eqnarray}
F_1 &=&  0 \nonumber\\ 
F_2 &=&  8 \pi^3 \,\frac{{\cal Q}^2}{x} (\Delta - 1) \,\Gamma (\Delta ) 
\Big[ \frac{q^2}{4c}\Big]^3 \nonumber\\
&\times & \!\!\! \frac{\Gamma \Big( \frac{q^2}{4c}\Big( \frac{1}{x} - 1 \Big)+ \Delta -1 \Big) } {\Gamma\Big(\frac{q^2}{4c}\Big( \frac{1}{x} - 1 \Big) + 1 \Big) } 
\Big[\frac{ \Gamma \Big( \frac{q^2}{4c\,x}   ) }
{\Gamma( \frac{q^2}{4c \, x} + \Delta  \Big) } \Big]^2 \,, \nonumber \\
\label{F2escalar}
\end{eqnarray}

\noindent which agrees at leading order in $ 4c/q^2 $  with the hard wall structure functions 
(\ref{hwsf}) as shown in \cite{BallonBayona:2007qr}.


\subsection{ DIS in D3-D7 system   }     

In the D3-D7 brane model one considers an $AdS_5\times S^5 $ space and includes $N_f$ coincident 
D7 probe branes. In order to represent the localization of the D7 branes, it is more convenient 
to write the $AdS_5\times S^5 $  metric in cylindrical coordinates\cite{BallonBayona:2008zi}:
\begin{eqnarray}
& ds^2_{10}& \!\!\!\! =  \frac{\rho^2+w_5^2 +w_6^2}{R^2}\eta_{\mu\nu}dx^\mu dx^\nu \nonumber\\
&+& \!\!\!\!\!\!\!\!\frac{R^2}{\rho^2+w_5^2 +w_6^2}\Big[ dw_5^2 + dw_6^2 +d\rho^2 + \rho^2 d\Omega_3^2 \Big] ,
\nonumber\\
\end{eqnarray}

In these new coordinates, the localization of the D7 branes can be chosen as $w_5 = 0 $, $ w_6 = L$. The metric induced on the brane is then
\begin{eqnarray}
\label{branemetric} 
&ds^2_8& \!\!\!\!  = G_{ab} dx^a dx^b \,=\,   \frac{\rho^2 + L^2}{R^2}  \eta_{\mu\nu} \,dx^\mu dx^\nu 
\nonumber\\
&+& \!\!\!\! \frac{R^2}{\rho^2 + L^2} \big( d\rho^2  + \rho^2 d\Omega_3^2, \big). 
\end{eqnarray}

\noindent Note that $\rho^2 + L^2 = r^2$ so that  the AdS radial coordinate $r$ on the brane 
is restricted to $ L  \le r < \infty $. This corresponds to an induced infrared cut off:
$m_h = L/R^2$.

In this model scalar mesons correspond to fluctuations of the D7 brane in the transversal directions 
$w_5$, $\,w_6$.  The solutions are written in terms of hypergeometric functions.  
The spectrum of states is given by $ - p_n^2 \,=\, 4 m_h^2 [ (n + \ell +1)(n+ \ell + 2) ] $, where
$\ell (\ell + 2) $ is the eigenvalue of the angular laplacian in $S^3$, related to the conformal dimension $\Delta\,$ of the scalar hadron by: $ \Delta = \ell + 3 \,$. 

The sum over the masses of the final states appearing in eq. (\ref{Imag}) for the D3-D7 model reads:
\begin{eqnarray}
\label{massasD3D7}
&\sum_{_X} & \!\!\!\! \delta \Big( M_X^2 + (P+q)^2 \, \Big) 
\,\approx \,  \frac{1}{4 m_h^2 ( 2n + 2\ell + 3)}\nonumber\\
&=& \!\!\!\! \frac{1}{4 m_h \sqrt{ s + m_h^2}}\,.
\end{eqnarray}

There are two possibilities for the gauge field to be taken as the approximate dual 
to the boundary hadronic current: a gauge field
living in the AdS bulk and a gauge field living on the D7 brane.

\bigskip

\noindent{\bf Case A: Gauge field in the AdS bulk}

This corresponds to the following interaction action
\begin{equation}
\int d^{8}x \sqrt{- G} \,
\frac{v^\alpha A^{\tilde m}}{\rho^2 + L^2} \Big(
\partial_\alpha \Phi^\ast  \partial_{\tilde m} \Phi  
+ \partial_{\tilde m} \Phi^\ast  \partial_{\alpha} \Phi \Big)
.
\label{acaoint1}
\end{equation}
   
For this case, the structure functions found in \cite{BallonBayona:2008zi} are
\begin{eqnarray}
\label{StructureD3D7}
F_1 &=&  0 \nonumber\\
F_2 &=&  8\pi^3 {\cal Q}^2 \, \frac{\Gamma(2\ell + 4)}{\Gamma^4 (\ell + 2)} \frac{\Gamma(n + 2\ell + 3)}{\Gamma (n + 1)} \nonumber\\
& \times & \!\!\!\!  \left(\frac{q^2}{m_h^2}\right)^2 \frac{{\cal I}^2_{bulk}}{x}\,.
\end{eqnarray}

\noindent where ${\cal I}_{bulk}$ is an integral with no analytical solution, that can be approximated near the elastic limit $\,x \to 1\,$, where we find for $ q^2 >> m_h^2 $  
\begin{eqnarray}
F_2 \sim \left(\frac{4{m_h}^2}{q^2}\right)^{\ell + 2 } (1-x)^{\ell + 1}
\nonumber\\
= \left(\frac{4{m_h}^2}{q^2}\right)^{\Delta-1} (1-x)^{\Delta-2}\,,
\label{F2hw}
\end{eqnarray}

\noindent which agrees with the hard and soft wall model results in this regime.


\bigskip

\noindent{\bf Case B: Gauge field on the D7 brane}

Gauge fields living on the D7 brane were studied in \cite{Kruczenski:2003be}, from the  Dirac-Born-Infeld action for the brane. The interaction between this gauge fields and the scalar fields that lives also on the D7 brane can be obtained by imposing gauge invariance 
\begin{equation}
S_{\Phi} \,=\,- \, \int d^{8}x \sqrt{- G} \frac{G^{ab}}{\rho^2 + L^2}
\nabla_a \Phi^\ast  \nabla_b \Phi  \,,
\label{acaobranacovariante}
\end{equation}

\noindent where $\nabla_a \,=\, \partial_a \,-\, i \frac{\cal Q}{R} A_a \,$. From this action we obtain the three point interaction term
\begin{eqnarray}
\langle  S_{int}  \rangle = i\,\frac{\cal Q}{R}\, \int d^{8}x \sqrt{- G} \,G^{\mu\nu} A_\mu  \frac{1}{\rho^2 + L^2}\nonumber\\
\times  \Big( \Phi_i   \partial_{\nu} \Phi^\ast_X \,-\, \Phi^\ast_X \partial_{\nu} \Phi_i \,  \Big)  \,,
\label{acaoint3}
\end{eqnarray}

\noindent Following again a prescription associating this interaction action with 
the matrix element of the hadronic current we found $F_1 = 0$ and an $F_2$ that 
can be approximated  in the regime $\, x<< 1\, $ by
\begin{eqnarray}
\label{Structurebrane2}
F_2  \approx  \pi^5 {\cal Q}^2  \frac {(1.27)^2}{2} \Gamma(2\ell + 4) (\ell+2)^2 
\nonumber\\
\times \left(\frac{ 4 m_h^2  }{q^2}\right)^{\ell+2}\, x^{\ell+4} \,.
\end{eqnarray}

\begin{figure}
\centering
\includegraphics[width=6cm]{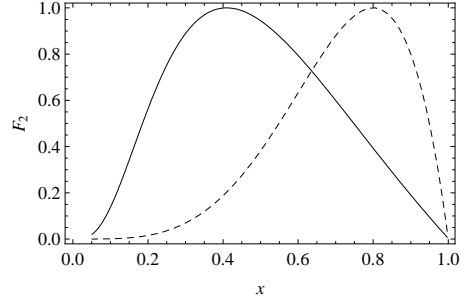}
\vspace{-1cm}
{\caption{ $F_2$ vs x for $q/(2m_h)=75$ and $\ell=0$. The dashed line correspond to the case A while the solid line correspond to the case B.}\label{F2vsX}}
\end{figure}

In Figure \ref{F2vsX} we plot our numerical results for the structure function $F_2$ as a function of $x$ for $q/(2m_h)=75$ and $\ell=0$ for the cases A and B. 

  

  
\bigskip

\noindent{\bf  Elastic form factors}

Hadronic elastic form factors can also be calculated from the previous results. 
The form factors are defined in terms of the matrix element of the hadronic current
in elastic regime. Explicitly, the form factor $F(q^2 )$ 
is defined by
\begin{equation}
\langle P + q, X \vert J^\mu ( 0 )  \vert P,{\cal Q} \rangle \,= \,2\, F(q^2)\,[2P^{\mu}+q^{\mu}].
\end{equation}  

For both cases of field in the bulk or on the brane we find
\begin{eqnarray}
\label{formfactor}
F(q^2)= \frac{Q}{2}\, \Gamma(2\Delta-2)(\Delta-1)\Big(\, \frac{4m_h^2}{q^2}\Big)^{\Delta-1},
\end{eqnarray}

\noindent in agreement with results from D3-D7 \cite{Hong:2003jm} and hard and soft wall models  
 \cite{Grigoryan:2007vg,Grigoryan:2007my,Brodsky:2007hb}.


\section{DIS when final states have higher conformal dimension} 

In the previous sections it was considered that the final hadronic states have the same conformal dimension as the initial ones. In ref. \cite{BallonBayona:2007rs} we calculated DIS structure functions summing over all allowed values of final conformal dimension$ \Delta' $:
\begin{equation}
F_2(x,q^2)\,=\,  F^{^{\Delta' = \Delta}}_2(x,q^2) \,+\, \sum_{\Delta' > \Delta } F^{^{\Delta' }}_2(x,q^2)  \,.
\end{equation}

The conformal dimension is related to the number of constituents of the hadron.
For a fixed value of the energy and momentum of the final state, the maximum number of hadrons that can be produced by the final state, assuming that all have a minimum mass $\Lambda$ is
\begin{equation}
\label{Nmax}
N_{max} \,\approx\, \frac{\sqrt{s}}{\Lambda} \,\approx\, \Big( \,\frac{q^2}{x \Lambda^2}\, \Big)^{1/2}\,.
\end{equation}
This places an upper bound in the sum over conformal dimensions and leads, in the small $x$ limit to 
\begin{equation}
\label{Final}
F_2 (x,q^2)\,\approx\, 
\pi C_0 \, {\cal Q}^2 \left( \frac{q^2}{\Lambda^2} 
\right)^{1/2} \, x^{-1/2} \,\,.
\end{equation}

This result shows a behavior similar to the observed geometric scaling \cite{Stasto:2000er} of 
total photon-hadron cross sections. The total cross section is related to the structure function $F_2 $ by 
\begin{equation}
\sigma (q^2 , x)  \, =\, 4 \pi^2 \alpha_{EM} \, \frac{ F_2 (x,q^2)}{q^2}.
\end{equation}

\noindent and geometric scaling means that this quantity depends on $q^2$ and $x$ only through the 
combination $ q^2 x^\lambda $ with $ 0.3 < \lambda < 0.4 $. Our result shows a similar scaling with 
$ \lambda = 1$. 

\bigskip 

\noindent{\bf Acknowledgments.} 
We would like to thank partial financial support from CNPq, Capes and Faperj, 
Brazilian agencies. 


\end{document}